%% file: main.tex
\documentclass[conference,a4paper]{APSIPA2021}
\usepackage{amsmath}
\usepackage{graphicx}
\usepackage{multirow}
\usepackage{threeparttable}
\usepackage[backend=biber,style=ieee,]{biblatex}
\usepackage{geometry}
\usepackage{url}
\usepackage{amsfonts}

\usepackage{subcaption}

\usepackage{makecell}
\usepackage{booktabs}

\usepackage{fancyhdr}

\fancypagestyle{firststyle}{
  \fancyhf{}
  \fancyhead[C]{2023 Asia Pacific Signal and Information Processing Association Annual Summit and Conference (APSIPA ASC)}
}

\begin{document}

\title{Evaluating Methods for Ground-Truth-Free Foreign Accent Conversion}

\author{
\authorblockN{
Wen-Chin Huang and Tomoki Toda
}

\authorblockA{
Nagoya University, Japan \\
E-mail: wen.chinhuang@g.sp.m.is.nagoya-u.ac.jp
}}

\maketitle
\thispagestyle{firststyle}
\pagestyle{fancy}

\begin{abstract}
Foreign accent conversion (FAC) is a special application of voice conversion (VC) which aims to convert the accented speech of a non-native speaker to a native-sounding speech with the same speaker identity. FAC is difficult since the native speech from the desired non-native speaker to be used as the training target is impossible to collect. In this work, we evaluate three recently proposed methods for ground-truth-free FAC, where all of them aim to harness the power of sequence-to-sequence (seq2seq) and non-parallel VC models to properly convert the accent and control the speaker identity. Our experimental evaluation results show that no single method was significantly better than the others in all evaluation axes, which is in contrast to conclusions drawn in previous studies. We also explain the effectiveness of these methods with the training input and output of the seq2seq model and examine the design choice of the non-parallel VC model, and show that intelligibility measures such as word error rates do not correlate well with subjective accentedness. Finally, our implementation is open-sourced to promote reproducible research and help future researchers improve upon the compared systems.
\end{abstract}

\section{Introduction}
\label{sec:intro}

Voice conversion (VC) is the task of converting between two types of speech without changing the linguistic content \cite{vc-survey, vc-survey-2021}. While speaker conversion has been the main focus of previous VC research, in this paper we aim to tackle a special application: foreign accent conversion (FAC)\footnote{Readers should note that the term ``accent conversion'' can be referred to many different tasks in the literature. While many have used this term to refer to the conversion between different accents or from native to accented speech, in this work we focus on the task of ``de-accenting''.} \cite{fac}. As illustrated in the right side of Figure~\ref{fig:training-conversion-phases}, given an accented speech utterance spoken by a non-native source speaker, FAC aims to generate a native-sounding version with the same speaker identity as the source speaker. Applications of FAC include computer-aided language learning \cite{fac, fac-for-call, golden-speaker-builder}  and entertainment such as movie dubbing \cite{fac-dubbing}.

At first sight, one might try to apply existing state-of-the-art deep learning-based VC methods, such as sequence-to-sequence (seq2seq) modeling \cite{VTN, VTN-TASLP} or non-parallel recognition-synthesis-based models \cite{VC-PPG}, to FAC. However, such direct application is considered infeasible because of the absence of the ground-truth training target: it is impossible to collect native speech from a non-native speaker. In fact, there are many other VC applications that face the same problem, such as dysarthric speech conversion \cite{dvc-vtn-vae, n2d-vc}, which aims to convert from disordered speech to healthy speech while preserving the speaker identity.

Researchers have proposed several methods for such a type of ground-truth-free\footnote{\cite{stg} used the term `reference-free`, but we found it confusing since a reference speaker is needed for training. Therefore we use ``ground-truth-free'' in the rest of the paper, where ``ground-truth'' refers to the ground-truth used as the training target.} VC task. They all share a main idea, which is to first collect a training corpus from the source non-native speaker and then collect the native counterpart from a native reference speaker with the same prompt set, as depicted in the left side of Figure~\ref{fig:training-conversion-phases}. Then, state-of-the-art VC methods for disentangling the speaker and content are designed to achieve FAC. However, these works were developed in parallel, and comparisons were often conducted on a system-to-system basis, leading to a lack of comparison and understanding of each method.

\input{figures/training-conversion-phases}

In this work, we aim to systematically evaluate three methods \cite{dvc-vtn-vae, stg, lsc} for FAC. Experiments were conducted in a unified setting using a shared database, model architecture, and vocoder. We conducted a subjective evaluation test assessing three different aspects (naturalness, speaker similarity, accentedness) of the synthesized samples, allowing us to compare the methods.
As we will show in our experimental evaluation section, we found that \textbf{no single method was significantly better than the others in all evaluation axes}, which is in contrast to conclusions drawn in previous studies \cite{lsc}. We also present results of an objective intelligibility measure which was used in previous studies \cite{stg}, and show that it might not correlate well to subjective accentedness. Finally, to promote reproducible FAC research, we open-source our implementation to help future researchers improve upon our system\footnote{\label{github}\url{https://github.com/unilight/seq2seq-vc}}.

\section{Related works and background knowledge}

\subsection{Non-native to native foreign accent conversion}

In the FAC literature, most works tried to utilize accent-independent features to decompose accent from voice identity. For instance, early attempts made use of articulatory trajectories (e.g., lips and tongue movements) \cite{fac-gmm, fac-unit-selection, fac-dnn} and vocal tract length normalization \cite{fac-vtln}. More recently, more simplified features such as phonetic posteriorgrams (PPGs) \cite{fac-ppg, fac-ppg2} and text \cite{fac-asr-tts} are combined with advanced deep neural network architectures, especially seq2seq VC models, whose ability to model segmental and prosody features simultaneously play a crucial role in FAC. However, only very few works have tried to address ground-truth-free FAC \cite{stg, lsc}. We believe this difficult yet practical setting is worth further investigation.

\subsection{Seq2seq VC modeling}

Seq2seq modeling learns the alignment between the source and target data sequences in a data-driven manner, equipping the model with the ability to generate outputs of various lengths and capture long-term dependencies.
This is considered particularly important in FAC, as duration and supra-segmental characteristics like $f_0$ play an important role in prosody conversion, which strongly affects the perception of foreign accents.
In addition, in this work, we apply the text-to-speech (TTS) pre-training technique proposed in \cite{VTN, VTN-TASLP} to increase the performance. However, a major drawback of seq2seq models is the requirement of parallel data, which is impractical to collect in FAC. Therefore, we resort to the help of another family of VC models.

\subsection{Non-parallel frame-based VC modeling}
\label{ssec:npvc}

Non-parallel VC relaxes the requirement of a parallel corpus, and one of the mainstream methods is autoencoder-style training. An encoder first extracts latent features to filter out certain attributes (e.g. speaker identity in speaker conversion), and then a decoder tries to generate the input speech by reconstructing the missing information. In contrast with seq2seq models, non-parallel models are frame-based, i.e. the duration and other supra-segmental attributes are not converted. It is therefore assumed that non-parallel frame-based models convert only global characteristics such as the speaker identity while maintaining local characteristics, such as pronunciation.

Although the extractor and the decoder can be jointly trained, as reported in \cite{vcc2020}, it is more effective to first train the extractor and then used the extracted features to train the decoder. The extractor is usually pre-trained using a large, multi-speaker dataset, and as a result an any-to-one (A2O) VC model is obtained as it generalizes well to unseen speakers. Among the many choices, the most widely used latent feature is the PPG \cite{VC-PPG} extracted from a supervisedly trained automatic speech recognition (ASR) model, as it preserves strong linguistic clues and serves as a strong speaker information bottleneck. Recently, self-supervised speech representations (S3Rs) are attractive in that they can be trained without labeled data and thus benefit from training on larger scales. Many have applied them to VC as an alternative \cite{s3prl-vc, s3prl-vc-journal}, and the best performing one is vq-wav2vec \cite{vq-wav2vec}. In this work, we evaluate the effectiveness of PPG and vq-wav2vec in the context of FAC.

\section{Evaluated methods}

In all three methods we evaluate, we assume two materials are prepared beforehand. First, as shown in Figure~\ref{fig:training-conversion-phases}, access to a parallel dataset between the non-native speaker and a reference native speaker is assumed. Second, the dataset of the native speaker is used to train a non-parallel frame-based model, which will then be fixed in all three methods. This process is illustrated in the top of Figure~\ref{fig:methods}. In the following subsections, the detailed procedures of all three methods are described. Figure~\ref{fig:methods} shows a complete illustration.

\subsection{Method 1: cascade}

The cascade method was originally proposed for dysarthric VC \cite{dvc-vtn-vae}, but as mentioned in Section~\ref{sec:intro}, since dysarthric VC and FAC share a common problem of lacking ground-truth training target, here we examine whether it could be applied to FAC.

In the cascade method, a seq2seq model is trained to map from the source non-native speech to that of the reference native speaker. During conversion, the source speech is first sent into the seq2seq model to get the first stage converted speech. Although the nativeness is improved, the speaker identity is unwantedly changed into that of the reference speaker. Therefore, the non-parallel VC model is then used to change the identity back to that of the native speaker, while maintaining the pronunciation.

\subsection{Method 2: synthetic target generation (STG)}

In STG \cite{stg}, the non-parallel VC model converts the training dataset of the native speaker such that the generated speech has the same nativeness of the input but with the speaker identity of the non-native speaker. We refer to this step as synthetic target generation. Then, the seq2seq model is trained using the non-native training set as the source and the synthetic native speech with the speaker identity of the same non-native speaker as the target. The conversion process is then as simple as using the seq2seq model to generate the de-accented speech with the identity of the non-native speaker.

\subsection{Method 3: latent space conversion (LSC)}

The LSC method \cite{lsc} first uses the latent feature extractor module of the non-parallel VC model to transfer the training datasets of the source non-native and target native speakers from the speech space to the latent space. Then, the seq2seq model is trained to map the source latent features to the target latent features. During conversion, the latent features of the source non-native speech are first extracted and transformed to their native counterpart using the seq2seq model. Finally, the decoder of the non-parallel VC model is used to inject the identity of the non-native speaker into the converted latent features in order to generate the final converted speech.

\input{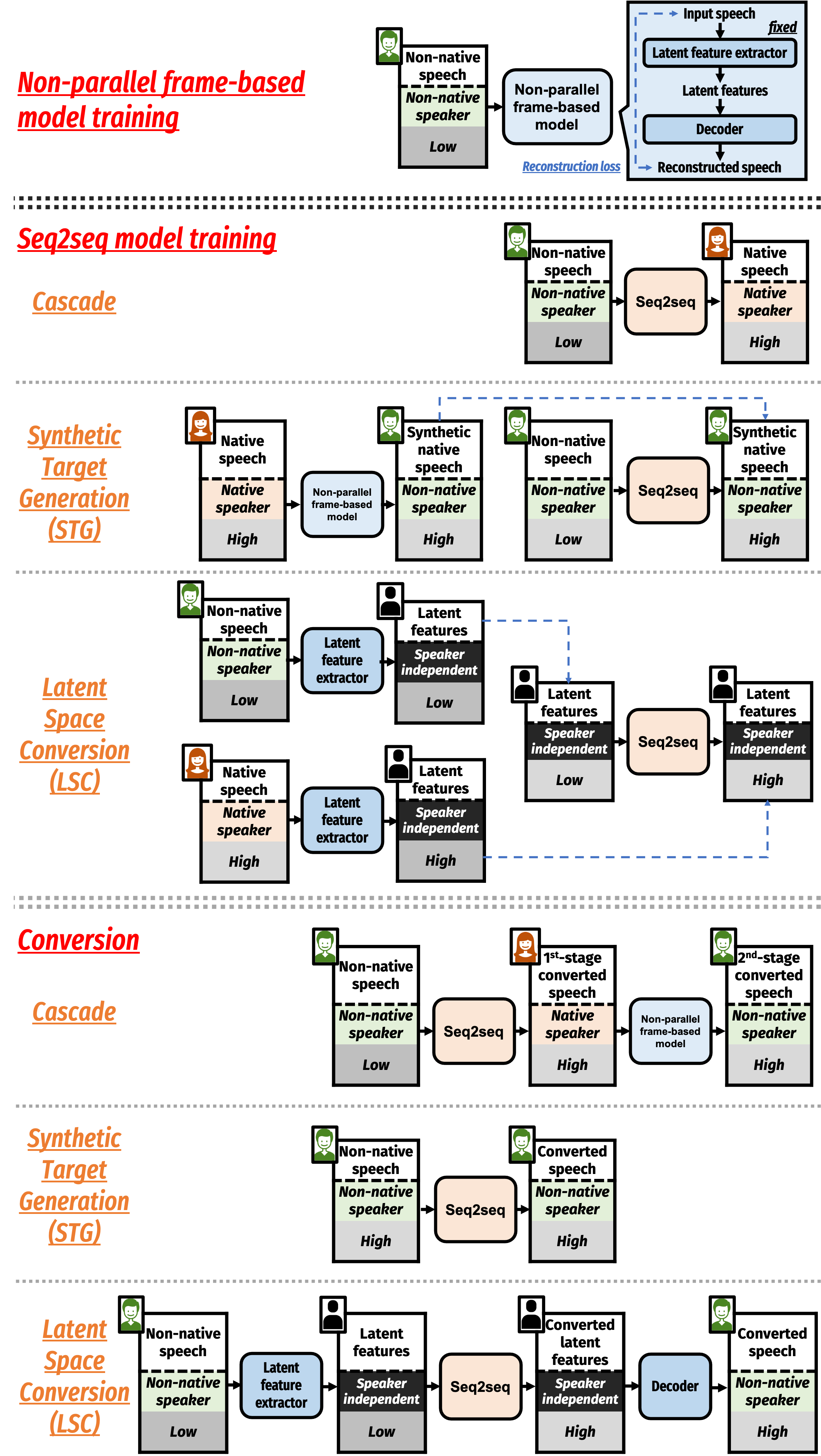}

\subsection{What is the difference between these three methods?}
\label{ssec:hypothesis}

As these methods might be seemingly complicated in their individual ways, the difference lies in the training input and output of the seq2seq model. This can be clearly observed in the right column of the middle part in Figure~\ref{fig:methods}. First, in the cascade method, the seq2seq model needs to simultaneously convert the speaker identity and the nativeness, which is considered to be the most difficult. In contrast, STG first changes the speaker identity of the native training set, such that the seq2seq model only needs to model the pronunciation pattern. Finally, to further ease the job of the seq2seq model, LSC projects the training datasets onto the latent space, which is speaker-independent, and this is easier to model than the speech space. Based on this observation, we may hypothesize that the performance of each method should correlate with the difficulty of the mapping function that the seq2seq model needs to find. We will discuss this hypothesis in the results section.

These three methods also have their own weakness. For instance, during the conversion phase, not only cascade but also LSC does the input speech pass through a pipeline consisting of multiple modules, suffering from potential error propagation. STG, on the other hand, requires only the seq2seq module during conversion and thus does not suffer from error propagation, but the synthetic target data inevitably contains artifacts. The performance of the seq2seq model is then bounded by how imperfect the synthetic data is. With these unique limitations, we note to the readers that it is difficult to fairly compare these methods.

\section{Experimental evaluation and discussion}
\label{sec:exp}

\subsection{Experimental setting}

The non-native and reference native speakers are THXC (Chinese male) and bdl (English male) from the L2-ARCTIC \cite{l2arctic} and ARCTIC datasets, respectively. There are 1032/50/50 training/development/testing parallel utterances respectively, and the total duration of the training set is around one hour. All samples are in 16 kHz. The downsampled LJSpeech dataset \cite{ljspeech17} was used for the seq2seq model pretraining. The LibriSpeech (960 hours) \cite{librispeech} and LibriLight \cite{librilight} (60k hours) datasets  were used to train the PPG and vq-wav2vec models, respectively.

Mel-spectrograms were used as the acoustic features with a hop size of 256. ParallelWaveGAN \cite{parallel-wavegan} was used as the neural vocoder, trained with only the training set of the source non-native speaker. The seq2seq model is based on the Transformer \cite{transformer, VTN} and the decoder of the non-parallel frame-based VC model resembles Tacotron2 \cite{Taco2}, following \cite{s3prl-vc}.
The implementation is open-sourced\footnote{See footnote \ref{github}}, so we refer readers to the source code for detailed hyperparameters due to space limits. An NVIDIA V100 GPU was used to train all the models, and the seq2seq model and non-parallel VC model in total took 6 hours to train.

\input{results}

\subsection{Evaluation protocols}

We conducted subjective evaluation tests on three axes, following previous works \cite{stg, lsc}. In the naturalness and accentedness tests, listeners were asked to rate the samples on a 5-point and a 9-point scale, respectively. Samples of the source non-native and target native speech were also included. In the similarity test, a source non-native sample and a converted sample were presented, and listeners were asked to judge whether the two samples were uttered by the same speaker on a four-point scale, following the same convention in voice conversion challenges \cite{vcc2020}. The naturalness and similarity tests were conducted using Amazon Mechanical Turk, and each of the 65 workers we recruited rated 20 samples (or sample pairs). As for the accentedness, due to the difficulty of the task itself, we conducted an in-lab study by recruiting 19 listeners to each listen to 40 samples. Audio samples are available for readers\footnote{\url{https://unilight.github.io/Publication-Demos/publications/fac-evaluate}}.

Finally, we consider one objective measure for calculating speech intelligibility: character/word error rates (CER/WER) obtained by running an ASR model on the speech samples. We used a  pretrained ASR model based on wav2vec 2.0\footnote{Performance and APIs can be found at \url{https://huggingface.co/facebook/wav2vec2-large-960h-lv60-self}} \cite{wav2vec2}.

\subsection{Design choice of the non-parallel frame-based model}

In Secion~\ref{ssec:npvc}, we mentioned we experimented with two types of latent features in the non-parallel frame-based model: PPG and vq-wav2vec. As shown in Table~\ref{tab:results}, in all three subjective evaluation axes (naturalness, similarity, and accentedness) and all three evaluated methods, using PPG was almost always significantly better than using vq-wav2vec. The only exception was that the naturalness scores were nearly identical when using vq-wav2vec and PPG in the LSC scenario.

Although previous works \cite{s3prl-vc, s3prl-vc-journal} have already shown that PPG outperforms vq-wav2vec in terms of naturalness and similarity, the superiority of PPG in accentedness implies the importance of linguistic supervision in the training of the latent extractor. In the rest of the section, we focus on the results of the three methods using PPG.

\subsection{Effectiveness of the three evaluated methods}

In this subsection, we try to compare the performance tendency with the hypothesis described in Section~\ref{ssec:hypothesis}.
We first look at the naturalness scores of the three methods in Table~\ref{tab:results}. As the confidence intervals overlap, there is no statistically significant difference between the three methods. This suggests that naturalness is not affected by the difficulty of the seq2seq mapping.

Next, for similarity, STG is significantly better than cascade and LSC, which again violates the above-mentioned hypothesis. Nonetheless, this result leads to two implications. First, as the seq2seq model in STG operates in the speech domain, the model prioritizes the learning of speaker identity generation over accent removal. Second, the relatively low similarity of LSC implies that the assumption of speaker independence of the latent features may be invalid in the context of FAC.

Finally, in terms of accentedness, the only significant difference that can be observed is the superiority of LSC over STG. Although this does not match the hypothesis, we note that the accentedness score of LSC is significantly better than cascade and STG when using vq-wav2vec. This suggests that LSC is more robust to the choice of the latent feature.

\subsection{Is character/word error rate a proper objective measure for FAC?}

The feasibility of using objective measures to predict subjective results is a long-standing problem in VC research \cite{vcc2020-prediction, voicemos2022}. Developing such a measure allows us to inspect the performance during system development without the expensive subjective evaluation process. Some previous works on FAC reported CER/WER as an indirect measure of accentedness, with the expectation that reducing accentedness can also reduce the error rates. Nonetheless, we try to investigate whether such a hypothesis is valid.

With the 8 data points in Table~\ref{tab:results}, the linear correlation coefficients between accentedness and CER/WER are 0.413 and 0.442, respectively. It can be then inferred that there is a weak yet insignificant correlation between accentedness and CER/WER. We thus conclude that there are other factors than intelligibility when it comes to accentedness, thus using CER/WER solely as an objective measure for FAC is unreliable.

\section{Conclusion and future works}

In this work, we systematically compared three methods for ground-truth-free FAC. Experiments were carried out in a unified setting, and subjective tests were conducted in terms of naturalness, speaker similarity, and accentedness. In addition to the detailed discussion of each method and evaluation axis presented in Section~\ref{sec:exp}, the most important message that the evaluation results show is that \textbf{no single method was significantly better than the other two in all evaluation axes}. While this may arise from the insufficiency of the evaluated methods, we also doubt if the evaluation protocols which we follow are proper. Below we list two directions that we wish to improve upon in the future.

\subsection{Adopting non-autoregressive (non-AR) seq2seq modeling}

We observed that the seq2seq VC models suffer from intelligibility issues, evidenced by the high error rates in Table~\ref{tab:results}. This mainly comes from the autoregressive (AR) design of the seq2seq models we adopted. With the advance of non-AR seq2seq modeling \cite{fastspeech, nar-seq2seq-vc} which is known for its robustness compared to AR models, it can be expected that adopting non-AR models in the FAC methods can further improve the performance.

\subsection{Better subjective evaluation protocols}

Although we adopted the absolute rating evaluation protocol following previous works \cite{stg, lsc}, the large confidence intervals observed in the naturalness and accentedness axes in Table~\ref{tab:results} suggest that a more reliable protocol needs to be developed. Listener feedback suggests that a 9-point scale test as used in \cite{munro1995foreign, stg, lsc} is too fine-grained to give precise ratings. Also, while it is easy to tell whether a sample is native or not, rating the degree of accentedness is rather difficult. Comparative measurements such as preference tests might be more suitable, as advised in \cite{compare-score}.

\subsection{More accurate accentedness evaluation}

Listeners mentioned that even as native English speakers, it is difficult to confidently rate accentedness. One way to improve this is to provide a training section containing utterances with different levels of accentedness, as the one provided in \cite{jasmin-cgn}. Another alternative is to directly recruit linguistics or educators, as someone with in-depth professional knowledge may make judgments more confidently.

\section*{Acknowledgment}
This work was partly supported by JSPS KAKENHI Grant Number 21J20920, JST CREST Grant Number JPMJCR19A3, and a project, JPNP20006, commissioned by NEDO, Japan.

\printbibliography

\end{document}

%% file: figures/training-conversion-phases.tex
\begin{figure}[t]
	\centering
	\includegraphics[width=\columnwidth]{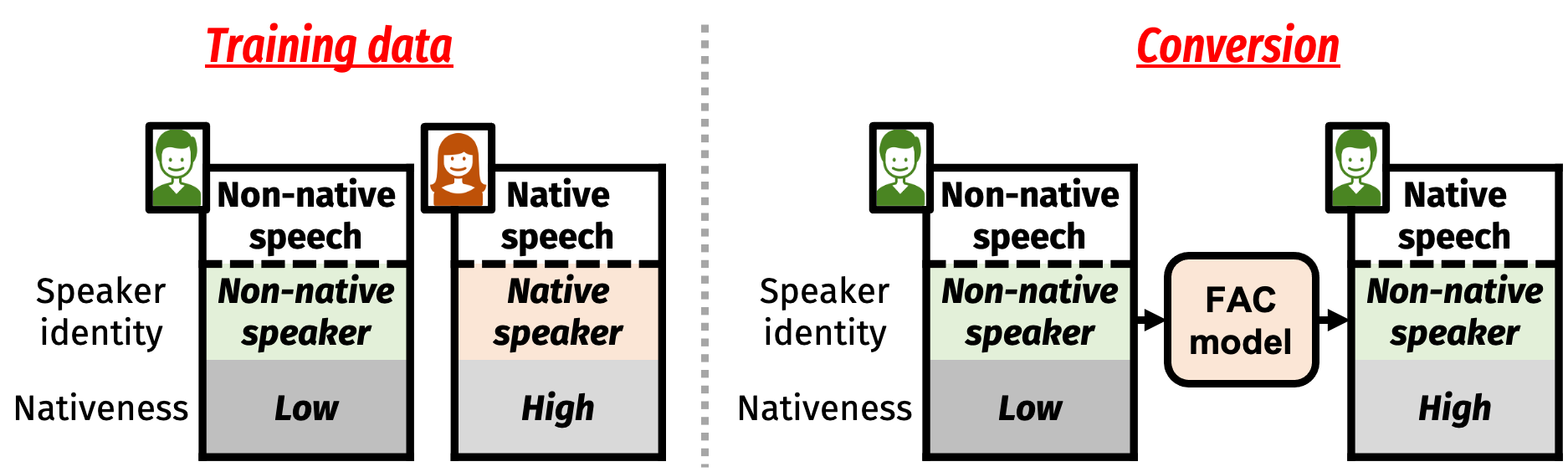}
	\centering
	\captionof{figure}{Left: the training data, which is a parallel corpus between the source non-native speaker and a reference native speaker. Right: the goal in the conversion phase of FAC. The nativeness is expected to be increased while maintaining the speaker identity.}
	\label{fig:training-conversion-phases}
     \vspace{-0.5cm}
\end{figure}

%% file: figures/training-conversion.tex
\begin{figure}[t]
	\centering
	\includegraphics[width=0.95\columnwidth]{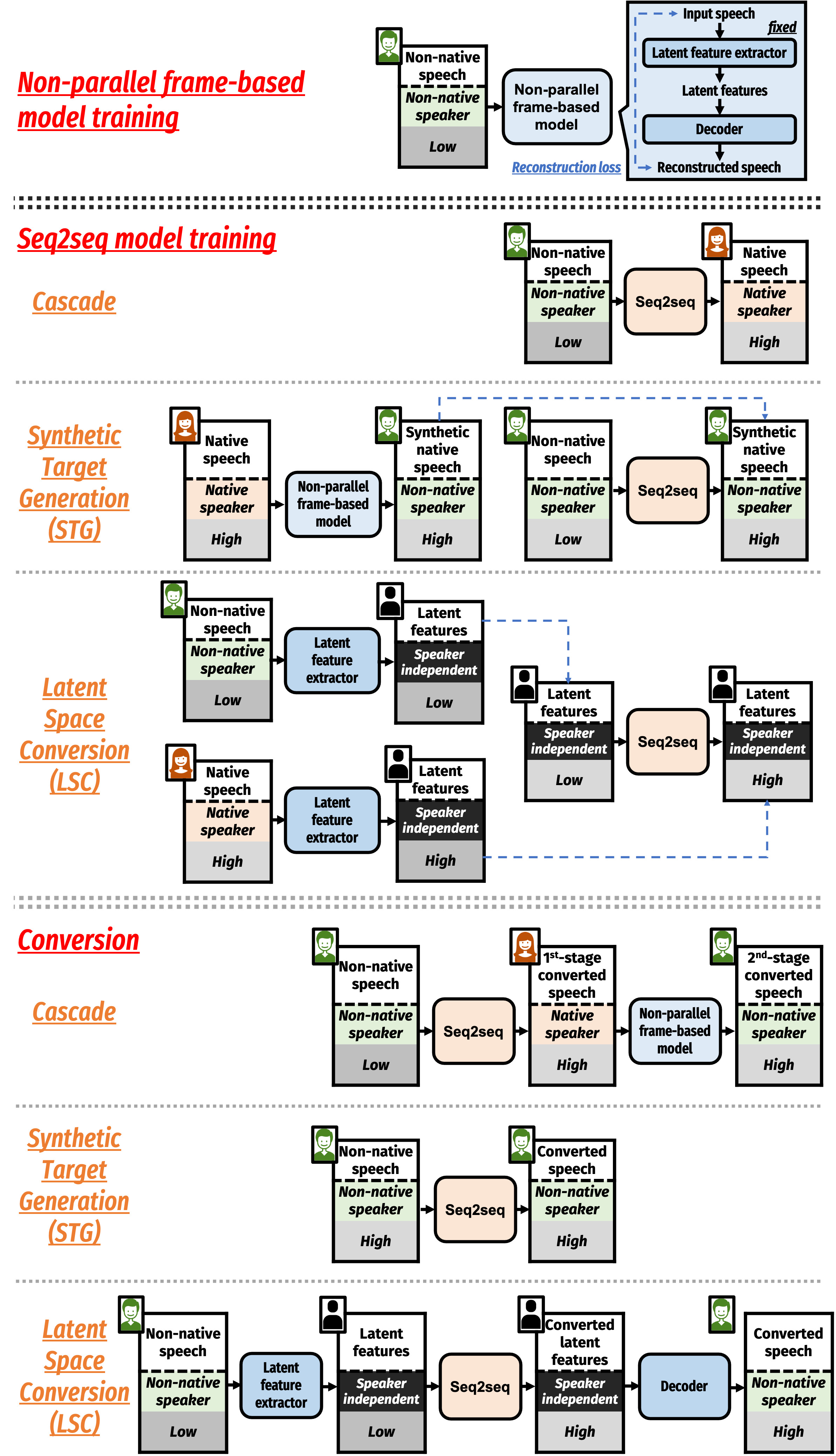}
	\centering
	\captionof{figure}{Illustration of the training and conversion processes of the three evaluated methods (cascade, STG, LSC). Top: non-parallel frame-based VC model training using the data from the source non-native speaker. Middle: seq2seq model training. All the components from the non-parallel frame-based VC model (marked in blue), including the latent feature extractor and the decoder, are fixed in this phase. Bottom: the conversion phases.}
	\label{fig:methods}
        \vspace{-0.5cm}
\end{figure}

%% file: results.tex
\begin{table*}[ht]
	\centering
	\caption{Objective and subjective evaluation results with 95\% confidence interval of samples from the evaluated methods, source and target.}
	
	\centering
        \begin{tabular}{ c c | c | c c c }
		\toprule
            Method & Extractor & CER/WER & \makecell{Naturalness $\uparrow$\\(1-5)} & \makecell{Similarity $\uparrow$ \\ (0\% -100\%)} & \makecell{Accentedness $\downarrow$\\(1-9)} \\
            \midrule
            \multicolumn{2}{c|}{Source (non-native)} & 5.3/12.3 & 4.18$\pm$0.19 & -- & 6.06$\pm$0.38 \\
            \midrule
		\multirow{2}{*}[0pt]{Cascade} & vq-wav2vec & 29.1/52.5 & 3.17$\pm$0.23 & 28.7\%$\pm$6.7\% & 5.41$\pm$0.32 \\
            & PPG & 30.4/52.7 & 3.50$\pm$0.22 & 45.7\%$\pm$7.3\% & 4.18$\pm$0.30 \\
            \multirow{2}{*}[0pt]{STG} & vq-wav2vec & 25.3/45.0 & 3.23$\pm$0.21 & 37.0\%$\pm$7.0\% & 5.27$\pm$0.31 \\
            & PPG & 17.7/40.9 & 3.66$\pm$0.20 & 57.3\%$\pm$7.8\% & 4.36$\pm$0.32 \\
            \multirow{2}{*}[0pt]{LSC} & vq-wav2vec & 33.4/52.5 & 3.65$\pm$0.25 & 36.0\%$\pm$7.0\% & 4.61$\pm$0.32 \\
            & PPG & 9.8/19.5 & 3.64$\pm$0.22 & 43.8\%$\pm$7.5\% & 3.95$\pm$0.31 \\
            \midrule
            \multicolumn{2}{c|}{Target (native)} & 1.3/4.3 & 4.42$\pm$0.18 & -- & 1.49$\pm$0.21 \\
		\bottomrule
	\end{tabular}

	\label{tab:results}
\end{table*}
